\documentclass[apjl]{emulateapj}
\usepackage{apjfonts}
\begin{document}
\shortauthors{Fish}
\shorttitle{$\Lambda$-Doubling Transitions of SiH in Orion KL}
\title{Search for $\Lambda$-Doubling Transitions of S\lowercase{i}H in
  Orion KL}
\author{Vincent L.~Fish\altaffilmark{1}}
\affil{National Radio Astronomy Observatory}
\affil{P. O. Box O, 1003 Lopezville Road, Socorro, NM  87801}
\email{vfish@nrao.edu}
\altaffiltext{1}{Jansky Fellow.}
\begin{abstract}
A recent submillimeter line survey of Orion KL claimed detection of
SiH.  This paper reports on GBT observations of the 5.7 GHz
$\Lambda$-doubling transitions of SiH in Orion.  Many recombination
lines, including C164$\delta$, are seen, but SiH is not detected.  The
nondetection corresponds to an upper limit of $1.5 \times
10^{15}$~cm$^{-2}$ ($4\,\sigma$) for the beam-averaged column density
of SiH.  This suggests that the fractional abundance of SiH in the
extended ridge is no more than twice that in the hot core.
\end{abstract}
\keywords{ISM: molecules --- ISM: individual (Orion KL) --- radio
  lines: ISM}

\section{Introduction}

Interstellar hydrides are the building blocks of common molecules.  In
dense regions, unsaturated hydrides represent a critical intermediate
step toward the formation of saturated hydrides as well as more
complex oxygenated and nitrogenated molecules.  For instance, SiO,
SiN, SiS, SiC, and other molecules may be produced in large part due
to gas-phase reactions with SiH \citep{turner77,mackay95}.  But
despite the important role that simple hydrides play in
astrochemistry, many have never been observed \citep{vandishoeck95}.
In the case of the SiH, this may be partly due to the fact that early
estimates of the $\Lambda$-doubling frequencies were highly uncertain.
The $\Lambda$-doubling frequency of 2940 MHz reported by
\citet{douglas65} has a large (10\%) uncertainty due to the
extrapolation of results obtained experimentally at high rotational
levels to low rotational levels.  Approximate calculations by
\citet{wilson75} for the ground-state triplet at 3.0 GHz are in error
by more than 150 MHz.  More recent calculations by Brown, Curl, \&
Evenson (1984,1985) give the transition frequencies to an accuracy of
3 MHz or $\sim 0.1$\%.

\begin{figure}[t]
\resizebox{\hsize}{!}{\includegraphics[]{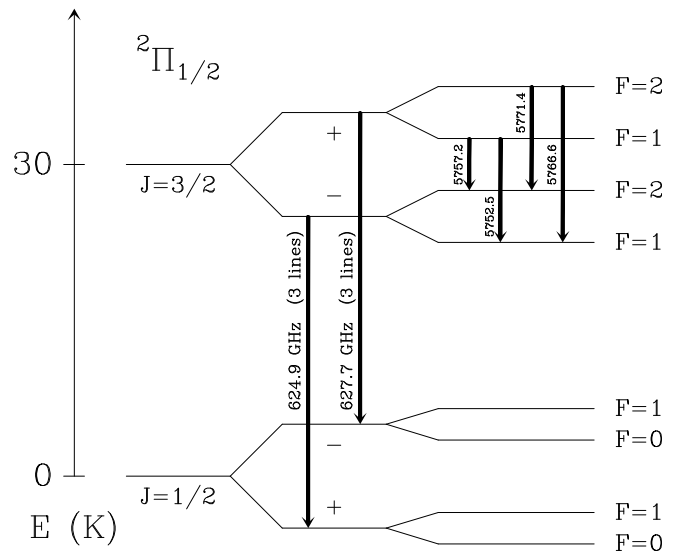}}
\caption{Diagram of the ground state and first rotationally excited
  state of SiH.  \citet{schilke01} claim a tentative detection of the
  two $^2\Pi_{1/2}, J = 3/2 \rightarrow 1/2$ triplets.  Observations
  of the indicated $J = 3/2$ $\Lambda$-doubling transitions are
  presented in this paper.  Frequencies are from
  \citet{brown85}.\label{fig-levels}}
\end{figure}

SiH (also known as silylidyne and silicon hydride) is isovalent to
the CH radical (methylidyne).  The ground electronic state is
$^2\Pi_{1/2}$, with a triplet of $\Lambda$-doubling $J = 1/2$
transitions near 3000 MHz \citep[e.g.,][]{brown85}.  This frequency is
inaccessible from most present radio telescopes.  However, the
first-excited state ($^2\Pi_{1/2}, J = 3/2$) $\Lambda$-doubling
quartet is found at 5.7 GHz (Figure \ref{fig-levels}), which is
tunable at many radio telescopes.

In a recent submillimeter line survey of Orion KL,
\citet{schilke01} report the first and only tentative detection of
interstellar SiH.  However, the six hyperfine $^2\Pi_{1/2}, J = 3/2
\rightarrow 1/2$ transitions are blended with strong emission from
more common molecules (SO$_2$, CH$_3$CN, and CH$_3$OCH$_3$).
\citeauthor{schilke01} regard their detection as merely tentative and
suggest that an interferometric follow-up may be required to determine
whether the detected features are due to SiH.

This paper reports on observations of the 5.7 GHz $\Lambda$-doubling
lines of SiH in order to attempt to confirm the \citeauthor{schilke01}
result in the centimeter wavelength regime.  Because the density of
molecular line transitions at centimeter wavelengths is much lower
than in the submillimeter, detection of the 5.7 GHz lines would
conclusively demonstrate that SiH is indeed present in Orion KL.  An
additional motivation was the possibility of being able to determine
the frequencies to much better accuracy than the 3 MHz obtained by the
laboratory measurements of \citet{brown85}.

\begin{figure}[t]
\resizebox{\hsize}{!}{\includegraphics[]{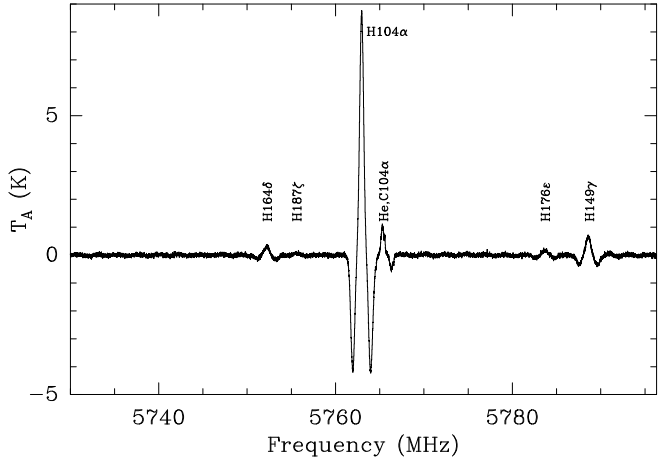}}
\caption{Spectral overview of Orion KL at 5760 MHz.  Calibrated
  frequency-switched data are shown with no baseline fitted.  Key
  recombination lines are identified.  These data were taken using a 1
  MHz switch frequency to clearly resolve the H104$\alpha$
  and He104$\alpha$ lines with minimal overlap.  Due to the method of
  frequency switching, approximately half-height negatives of lines
  appear offset by plus and minus the switch
  frequency.\label{fig-overview}}
\end{figure}

\begin{figure}
\resizebox{\hsize}{!}{\includegraphics[]{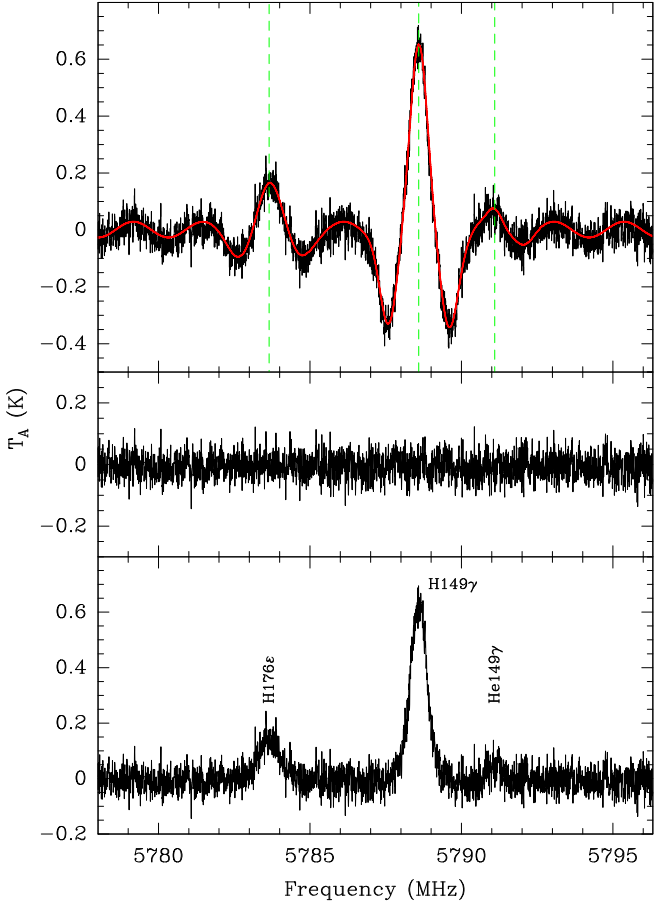}}
\caption{Enlargement of portion of Figure \ref{fig-overview} showing
  H149$\gamma$, He149$\gamma$, and H176$\epsilon$ features.
  \emph{Top}: Calibrated frequency-switched, Hanning-smoothed data
  using a switch frequency of 1 MHz.  The dashed green lines indicate the
  center frequencies of fitted Gaussian features.  A sinusoidal ripple
  has also been subtracted.  The red curve shows the sum of the
  Gaussian and sinusoidal fits.  \emph{Middle}: Fit residuals.
  \emph{Bottom}: Restored spectrum.  The positive Gaussian from each
  line fit is added back to the residuals.
  \label{fig-hi}}
\end{figure}

\section{Observations}

Data were taken in nine sessions from 2006 February 2 to 2006 March 2
with the Robert C.\ Byrd Green Bank Telescope (GBT), operated by
NRAO\footnote{The National Radio Astronomy Observatory is a facility
of the National Science Foundation operated under cooperative
agreement by Associated Universities, Inc.}.  The telescope was
pointed at Orion KL ($\alpha = 05^\mathrm{h}35^\mathrm{m}14\fs5,
\delta = -05\degr22\arcmin30\arcsec$, J2000).  The large beam
($2\farcm2$ FWHM) encompasses nearly 150~K of continuum emission,
precluding the use of position switching for calibration.  Frequency
switching with a switch frequency of 2~MHz was employed.

The GBT Spectrometer was used in 9-level mode to observe dual linear
polarization in each of four 12.5~MHz spectral windows centered at
5752.5, 5757.2, 5766.6, and 5771.4~MHz (the frequencies quoted by
\citealt{brown84,brown85}) in the frame $v_\mathrm{LSR} =
4.7$~km\,s$^{-1}$ radio, the velocity of the tentative SiH detection by
\citet{schilke01}.  Each spectral window was divided into 8192
spectral channels, providing a channel separation of 1.526~kHz
(0.08~km\,s$^{-1}$) before Hanning weighting.  In addition, some data
were taken with 50 MHz bandwidths using a switching frequency of 1
MHz.

Total time on source was approximately 32 hours.  A narrow
interference feature near 5760.8 MHz was seen during the first two
days of observations in the second spectral window.  Channels
containing this feature were flagged, as were the corresponding
channels offset by $\pm 2$~MHz.  No other interference was noted.

Data reduction was performed in GBTIDL.  A zeroth-order baseline was
subtracted from the data.  Spectral lines were fit with a Gaussian at
the center frequency and two negative Gaussians constrained to be at
the center frequency plus and minus the switch frequency with the same
linewidth.  The amplitude of the negative components was fit as a free
parameter; in the limit wherein $T_\mathrm{line} \ll T_\mathrm{sys}$,
the amplitude of the negative components approaches $-0.5$ times the
amplitude of the positive component.  In some IFs it was also
necessary to fit a sinusoid to remove baseline ripples.

\section{Results}

\subsection{Recombination Lines}

Figure \ref{fig-overview} shows an overview of the spectrum of Orion
KL around 5760 MHz based on 50 minutes of data using a 1 MHz switch
frequency.  The H104$\alpha$ recombination line dominates the
spectrum, with a series of hydrogen recombination lines visible down
to $\Delta n = 6$.  He104$\alpha$ and C104$\alpha$ lines are also
detected.  Figure \ref{fig-hi} contains an enlargement of the
high-frequency data.  The H149$\gamma$ and H176$\epsilon$ features are
clearly visible, and the He149$\gamma$ line is detected as well.

\begin{figure}[t]
\resizebox{\hsize}{!}{\includegraphics[]{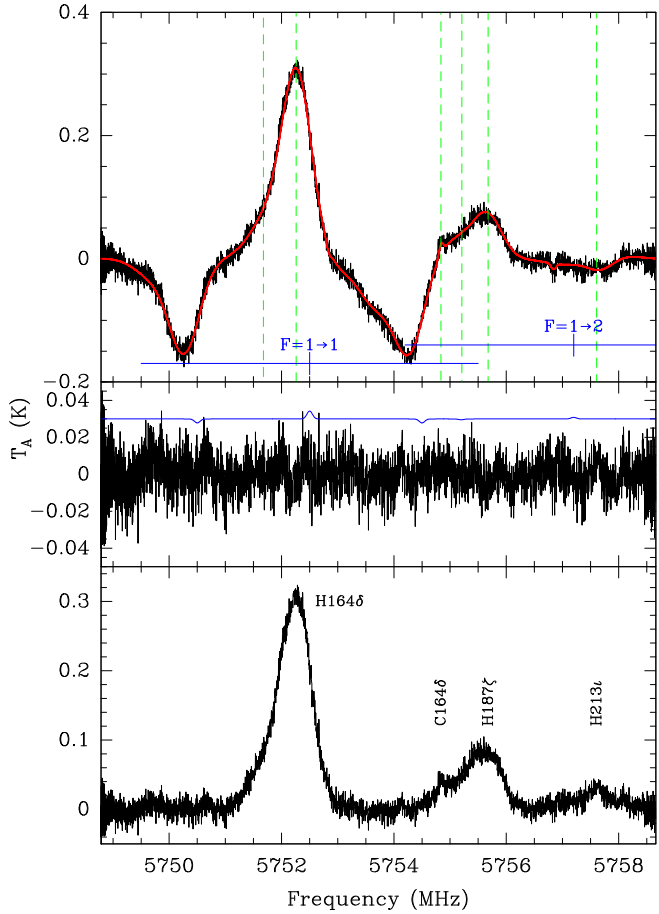}}
\caption{Spectral region of $F = 1 \rightarrow 1$ and $1 \rightarrow
  2$ lines.  A 2 MHz switch frequency was used.  Panels are as in
  Figure \ref{fig-hi}.  Blue lines indicate the calculated frequencies
  and ranges from \citet{brown85} as well as the expected signal from
  a beam-averaged column density of $1.5 \times 10^{15}$~cm$^{-2}$
  (see \S \ref{SiH-results}), vertically shifted for clarity.
  \label{fig-if0}}
\end{figure}

\begin{figure}[t]
\resizebox{\hsize}{!}{\includegraphics[]{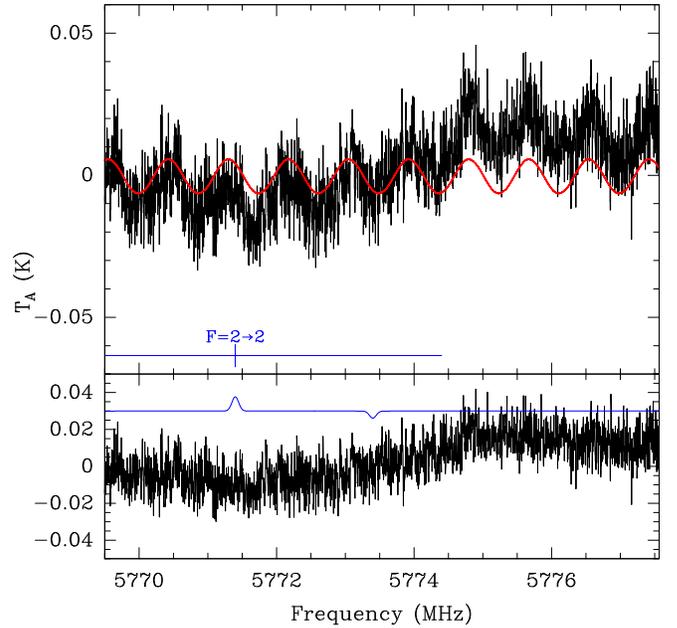}}
\caption{Spectral region of $F = 2 \rightarrow 2$ main line.  Two
  sinusoidal ripples have been fitted.  Some baseline structure
  remains, but no spectral lines are seen to within the noise limits.
  The top two panels are as in Figure \ref{fig-if0}.
  \label{fig-if1}}
\end{figure}

The spectral region near 5752.5 MHz (SiH, $^2\Pi_{1/2}, J=3/2, F = 1
\rightarrow 1$) is shown in Figure \ref{fig-if0}.  Several
recombination lines are seen, most prominently H164$\delta$.  Line fit
parameters and line identifications are presented in Table
\ref{tab-results}.  A two-component Gaussian fit is used for each of
the H164$\delta$ and H187$\zeta$ lines.  The lower-frequency component
of the H164$\delta$ line corresponds to a rather high velocity, but
the linewidth suggests that this component is not molecular in origin.
The lower-frequency component of the H187$\zeta$ line is poorly
determined, most likely due to blending with He164$\delta$.  The
H213$\iota$ line is also poorly fit due to its approximately 2 MHz
offset from the H187$\zeta$ line; any error in fitting the latter will
result in an incorrect fit of the former.  The quality of these fits
may also be affected by possible baseline errors associated with the
lack of line-free channels in the frequency-switched spectrum.  No
features are seen in the spectral window centered at the other main
line ($F = 2 \rightarrow 2$) frequency, 5771.4 MHz (Figure
\ref{fig-if1}).

\begin{figure}[t]
\resizebox{\hsize}{!}{\includegraphics[]{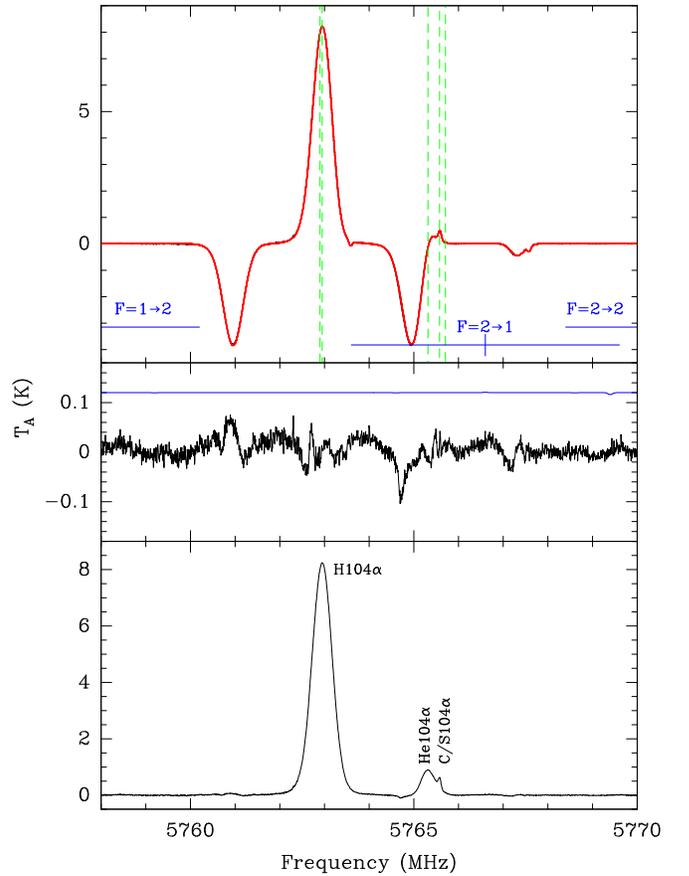}}
\caption{Spectral region of $F = 2 \rightarrow 1$ line.  Small errors
  in fitting the H104$\alpha$ feature do not strongly affect the
  quality of fits of He104$\alpha$ and C 104$\alpha$.  Panels are as
  in Figure \ref{fig-if0}.
  \label{fig-ifs23}}
\end{figure}

The spectral region containing the frequencies of the $F = 1
\rightarrow 2$ and $2 \rightarrow 1$ lines is shown in Figure
\ref{fig-ifs23}.  The figure shows the combined data from two IFs,
since neither IF spanned a large enough range of frequency to cover
both the bright H104$\alpha$ line and its two negatives at plus and
minus the switching frequency.  The data from one IF have been scaled
to produce a continuous spectrum in the region of overlap.  In
addition to the H104$\alpha$ line, He104$\alpha$, C104$\alpha$ and a
heavier element 104$\alpha$ recombination line are seen.  Small ($\sim
1\%$ of peak) fitting errors exist in the vicinity of the H104$\alpha$
line and its negatives, most likely due to the inadequacy of a
two-component Gaussian model to fit the spectrum.  Despite the overlap
of the high-frequency negative component with the other 104$\alpha$
recombination lines, the fits to the latter are not strongly affected
owing to the clean negative version at 5767 MHz.  It is possible that
there are minor baseline ripples as well, but the paucity of line-free
bandwidth in this region makes it difficult to determine the baseline
structure to high accuracy.

\begin{deluxetable}{llrll}
\tabletypesize{\small}
\tablecaption{Line Fit Parameters\label{tab-results}}
\tablehead{
  \colhead{Frequency} &
  \colhead{} &
  \colhead{$v_\mathrm{LSR}$} &
  \colhead{FWHM} &
  \colhead{$T_\mathrm{A}$} \\
  \colhead{(MHz)} &
  \colhead{Line} &
  \colhead{(km\,s$^{-1}$)} &
  \colhead{(km\,s$^{-1}$)} &
  \colhead{(K)\tablenotemark{a}}
}
\startdata
$5751.678 \pm 0.034$ & H164$\delta$ (?)             &  $24.7 \pm  1.8$ & $36.1 \pm  2.2$ & $0.06$ \\
$5752.260 \pm 0.005$ & H164$\delta$                 &  $-5.6 \pm  0.3$ & $32.9 \pm  0.4$ & $0.30$ \\
$5754.836 \pm 0.006$ & C164$\delta$                 &  $ 9.7 \pm  0.3$ & $ 5.0 \pm  0.9$ & $0.02$ \\
$5755.210 \pm 0.238$ & H187$\zeta$\tablenotemark{b} &  $10.0 \pm 12.4$ & $46.6 \pm 13.0$ & $0.04$ \\
$5755.678 \pm 0.029$ & H187$\zeta$                  & $-14.4 \pm  1.5$ & $31.5 \pm  2.9$ & $0.07$ \\
$5757.608 \pm 0.020$ & H213$\iota$ (?)              & $-25.5 \pm  1.1$ & $46.0 \pm  2.5$ & $0.03$ \\ 
$5762.902 \pm 0.004$ & H104$\alpha$                 &  $-1.1 \pm  0.2$ & $44.7\tablenotemark{c}$ & $0.62$ \\
$5762.952 \pm 0.001$ & H104$\alpha$                 &  $-3.7 \pm  0.0$ & $26.8 \pm  0.0$ & $7.62$ \\
$5765.319 \pm 0.001$ & He104$\alpha$                &  $-4.7 \pm  0.0$ & $17.8 \pm  0.1$ & $0.91$ \\
$5765.580 \pm 0.001$ & C104$\alpha$                 &  $ 9.1 \pm  0.0$ & $ 4.7 \pm  0.1$ & $0.42$ \\
$5765.709 \pm 0.007$ & S104$\alpha$                 &  $10.9 \pm  0.3$ & $ 5.5 \pm  0.8$ & $0.05$ \\
$5783.651 \pm 0.010$ & H176$\epsilon$               &  $-5.4 \pm  0.5$ & $34.4 \pm  1.9$ & $0.14$ \\
$5788.588 \pm 0.002$ & H149$\gamma$                 &  $-4.1 \pm  0.1$ & $44.7 \pm  0.4$ & $0.63$ \\
$5791.098 \pm 0.023$ & He149$\gamma$ (?)            & $-12.0 \pm  1.2$ & $29.6 \pm  3.6$ & $0.06$ \\
\enddata
\tablenotetext{a}{Systematic errors are 10 to 20\%; random errors are
  usually significantly smaller.}
\tablenotetext{b}{Poor fit; likely blended with He164$\delta$.}
\tablenotetext{c}{Poorly constrained.}
\end{deluxetable}

\subsection{Comparison with Previous Work}

Orion A is a complicated star-forming complex.  Observations of other
H$\alpha$ lines find many broad components with LSR velocities between
$-6$ and $+1$~km\,s$^{-1}$ within our beam
\citep[e.g.,][]{pauls77,pankonin79}.  The detected H104$\alpha$
profile likely contains contributions from each of these components.

The velocity and linewidth of the He104$\alpha$ line are in excellent
agreement with values obtained for the He91$\alpha$ line by Natta,
Walmsley, \& Tielens (1994).  The parameters for the He149$\gamma$ and
He164$\delta$ lines are less well determined.  The former is based on
less than an hour of observing time, while the latter is blended with
stronger emission from H187$\zeta$.

The C104$\alpha$ line is detected at $v_\mathrm{LSR} =
+9.1$~km\,s$^{-1}$ and the C164$\delta$ line is detected at
$+9.7$~km\,s$^{-1}$ with linewidths of around 5~km\,s$^{-1}$.  This is
the first detection in emission of a carbon $\delta$ line, although
\citet{stepkin06} report on low-frequency carbon $\delta$ absorption.
These parameters are in excellent agreement with other detections of
carbon recombination lines in the Orion KL region
\citep[e.g.,][]{balick74,ahmad76,boughton78,natta94}. In addition, the
recombination line from a heavier element is detected.  The linewidth
of $5.5 \pm 0.8$~km\,s$^{-1}$ is consistent with the $3.5 \pm
3.3$~km\,s$^{-1}$ obtained by \citet{ahmad76} in the 85$\alpha$
series, and the velocity shift of $-6.7 \pm 0.3$~km\,s$^{-1}$ with
respect to the C104$\alpha$ line is consistent with the $-6.8 \pm
1.4$~km\,s$^{-1}$ from \citeauthor{ahmad76}.  Based on expected
element depletion, the most likely heavier element to produce a
recombination line is sulfur \citep[e.g.,][]{pankonin77,qaiyum83}.
The rest-frequency shift of S104$\alpha$ relative to C104$\alpha$ is
$-8.5$~km\,s$^{-1}$, but it is not unreasonable to expect the sulfur
line to be centered at a different velocity, due both to possible
blending with other lines (such as Mg and Si) and the fact that the
sulfur and carbon trace slightly different (though overlapping)
regions \citep{sternberg95}.

\subsection{SiH}
\label{SiH-results}

From the submillimeter data of \citet{schilke01}, the FWHM linewidth
of SiH is approximately 6~km\,s$^{-1}$, although the lines are blended
with stronger features.  Assuming that the $\Lambda$-doubling lines at
5.7 GHz have similar characteristics, the SiH lines would appear as
narrow features (like carbon recombination lines) rather than broad
features (like hydrogen recombination lines).  The only narrow lines
detected are identifiable as C104$\alpha$, S104$\alpha$, and
C164$\delta$.

The strongest constraints on SiH abundance come from the main lines.
The relative intensities of the four lines in local thermodynamic
equilibrium are 5:1:1:9 for the 5752.5 ($F = 1 \rightarrow 1$), 5757.2
($1 \rightarrow 2$), 5766.6 ($2 \rightarrow 1$), and 5771.4 MHz ($2
\rightarrow 2$) transitions, respectively \citep{brown85}.  Taking a
dipole moment for SiH of 0.124~D \citep{lewerenz83}, the Einstein A
coefficient for the $F = 2 \rightarrow 2$ transition is $2.1 \times
10^{-12}$~s$^{-1}$.  The nondetection of a feature in the 5771.4 MHz
spectral window places a $4\,\sigma$ limit on the strength of the line
at 7.7~mK for a 6~km\,s$^{-1}$ FWHM linewidth in emission.  (The $F =
1 \rightarrow 1$ main line is expected to be weaker and may be blended
with the strong H164$\delta$ emission.)  This places a beam-averaged
upper limit of $1.5 \times 10^{15}$~cm$^{-2}$ on the column density in
the upper level.  In this same level, \citet{schilke01} obtain an
estimate of $4.2 \times 10^{15}$~cm$^{-2}$ in a 12\arcsec\ beam.
Assuming that their detection is real, this suggests that the column
density of SiH is enhanced by at least a factor of 3 in the hot core
compared to the surrounding region.

The upper limit on the beam-averaged column density of SiH can also be
used to deduce an upper limit of its enhancement in the hot core
compared to the extended ridge, which fills approximately half the
beam of the present observations \citep[e.g.,][]{ungerechts97}.  The
column density of H$_2$ in the hot core is approximately a factor of 3
higher than in the extended ridge \citep[e.g.,][]{blake87}.  Thus, the
fractional abundance of SiH in the ridge is no more than twice that in
the hot core.

\section{Conclusions}

A search for the 5.7 GHz $\Lambda$-doubling lines of SiH in Orion KL
did not yield a detection.  The upper limit column density of $1.5
\times 10^{15}$~cm$^{-2}$ in the upper level suggests that the
fractional abundance of SiH is not significantly higher in the
extended ridge than in the hot core.  Numerous recombination lines
were detected, including several from helium, carbon, and possibly
sulfur.  Recombination line parameters are consistent with previous
observations in the literature.

\acknowledgements

The author wishes to thank T.\ Minter for assistance in setting up the
observations and helpful advice on reducing the data as well as D.~A.\
Roshi for the reference regarding carbon $\delta$ detection.

{\it Facility: GBT}


\begin{thebibliography}{}

\bibitem[Ahmad(1976)]{ahmad76} Ahmad, I.~A.\ 1976, \apj, 205, 379
\bibitem[Balick, Gammon, \& Doherty(1974)]{balick74} Balick, B.,
  Gammon, R.~H., \& Doherty, L.~H.\ 1974, \apj, 188, 45
\bibitem[Blake et al.(1987)]{blake87} Blake, G.~A., Sutton, E.~C.,
  Masson, C.~R., \& Phillips, T.~G.\ 1987, \apj, 315, 621
\bibitem[Boughton(1978)]{boughton78} Boughton, W.~L.\ 1978, \apj, 222,
  517 
\bibitem[Brown et al.(1984)]{brown84} Brown, J.~M., Curl,
  R.~F., \& Evenson, K.~M.\ 1984, \jcp, 81, 2884
\bibitem[Brown et al.(1985)]{brown85} Brown, J.~M., Curl,
  R.~F., \& Evenson, K.~M.\ 1985, \apj, 292, 188
\bibitem[Douglas \& Elliot(1965)]{douglas65} Douglas, R.~E., \&
  Elliot, G.~A.\ 1965, Can.~J.~Phys., 43, 496
\bibitem[Lewerenz et al.(1983)]{lewerenz83} Lewerenz, M., Bruna,
  P.~J., Peyerimholf, S.~D., \& Brienker, R.~J.\ 1983, Molec.\ Phys.,
  49, 1
\bibitem[Mackay(1995)]{mackay95} Mackay, D.~D.~S.\ 1995, \mnras, 274,
  694 
\bibitem[Natta et al.(1994)]{natta94} Natta, A.,
  Walmsley, C.~M., \& Tielens, A.~G.~G.~M.\ 1994, \apj, 428, 209
\bibitem[Pankonin, Walmsley, \& Harwit(1979)]{pankonin79} Pankonin,
  V., Walmsley, C.~M., \& Harwit, M.\ 1979, \aap, 75, 34
\bibitem[Pankonin et al.(1977)]{pankonin77} Pankonin, V., Walmsley,
  C.~M., Wilson, T.~L., \& Thomasson, P.\ 1977, \aap, 57, 341
\bibitem[Pauls \& Wilson(1977)]{pauls77} Pauls, T., \& Wilson, T.~L.\
  1977, \aap, 60, L31
\bibitem[Qaiyum \& Razaullah Ansari(1983)]{qaiyum83} Qaiyum, A., \&
  Razaullah Ansari, S.~M.\ 1983, \mnras, 205, 719
\bibitem[Schilke et al.(2001)]{schilke01} Schilke, P., Benford, D.~J.,
  Hunter, T.~R., Lis, D.~C., \& Phillips, T.~G.\ 2001, \apjs, 132, 281
\bibitem[Stepkin et al.(2006)]{stepkin06} Stepkin, S., Konovalenko,
  A., Kantharia, N., \& Shankar, U.\ 2006, \mnras, submitted
\bibitem[Sternberg \& Dalgarno(1995)]{sternberg95} Sternberg, A., \&
  Dalgarno, A.\ 1995, \apjs, 99, 565
\bibitem[Turner \& Dalgarno(1977)]{turner77} Turner, J.~L., \&
  Dalgarno, A.\ 1977, \apj, 213, 386
\bibitem[Ungerechts et al.(1997)]{ungerechts97} Ungerechts, H.,
  Bergin, E.~A., Goldsmith, P.~F., Irvine, W.~M., Schloerb, F.~P., \&
  Snell, R.~L.\ 1997, \apj, 482, 245
\bibitem[van Dishoeck(1995)]{vandishoeck95} van Dishoeck, E.~F.\ 1995,
  LNP Vol.\ 459: The Physics and Chemistry of Interstellar Molecular
  Clouds, 459, 225
\bibitem[Wilson \& Richards(1975)]{wilson75} Wilson, I.~D.~L., \&
  Richards, W.~G.\ 1975, \nat, 258, 133

\end{thebibliography}
\end{document}